\documentclass[prd,twocolumn,nofootinbib,aps,tightenlines,preprintnumbers,notitlepage,longbibliography,superscriptaddress]{revtex4-1}
\usepackage{graphicx}
\usepackage{color}
\usepackage{amsmath}
\usepackage{amsfonts}
\usepackage[colorlinks=true,citecolor=blue,urlcolor=blue]{hyperref}

\begin{document}
\title{Tighter Limits on Dark Matter Explanations of the Anomalous EDGES 21cm Signal}
\author{Ely D. Kovetz}
\affiliation{Department of Physics \& Astronomy, Johns Hopkins University, Baltimore, MD 21218, USA}
\author{Vivian Poulin}
\affiliation{Department of Physics \& Astronomy, Johns Hopkins University, Baltimore, MD 21218, USA}
\author{Vera Gluscevic}
\affiliation{School of Natural Sciences, Institute for Advanced Study, Einstein Drive, Princeton, NJ 08540, USA}
\affiliation{Department of Physics, University of Florida, Gainesville, Florida 32611, USA}
\author{Kimberly K. Boddy}
\affiliation{Department of Physics \& Astronomy, Johns Hopkins University, Baltimore, MD 21218, USA}
\author{Rennan Barkana}
\affiliation{Raymond and Beverly Sackler School of Physics and Astronomy, Tel-Aviv University, Tel-Aviv 69978, Israel}
\author{Marc Kamionkowski}
\affiliation{Department of Physics \& Astronomy, Johns Hopkins University, Baltimore, MD 21218, USA}

\begin{abstract}

We investigate the hypothesis that Coulomb-type interactions between dark matter (DM) and baryons explain the anomalously low 21cm brightness-temperature minimum at redshift $z\sim17$ that was
recently measured by the EDGES experiment.
In particular, we reassess the validity of the scenario where a small fraction of the total DM is millicharged, focusing on newly derived constraints from Planck 2015 cosmic microwave background (CMB) data. 
Crucially, the CMB power spectrum is sensitive to DM--baryon scattering if the fraction of interacting DM is larger than (or comparable to) the fractional uncertainty in the baryon energy density.
Meanwhile, there is a mass-dependent lower limit on the fraction for which the 
required interaction to cool the baryons sufficiently is so strong that it drives the interacting-DM temperature to the baryon temperature prior to their decoupling from the CMB. If this occurs as early as recombination, the cooling saturates. 
We precisely determine the viable parameter space for millicharged DM, and find that only a fraction $\left(m_\chi/{\rm MeV}\right)0.0115\%\lesssim f\lesssim 0.4\%$ of the entire DM content, and only for DM-particle masses between $0.5\,{\rm MeV}-35\,{\rm MeV}$, can be charged at the level needed to marginally explain the anomaly, without violating limits from SLAC, CMB, Big-Bang nucleosynthesis (BBN), or stellar and SN1987A cooling. 
In reality, though, we demonstrate that at least moderate fine tuning is required to both agree with the measured absorption profile and overcome various astrophysical sources of heating. 
Finally, we point out that a $\sim\!0.4\%$ millicharged DM component which is tightly coupled to the baryons at recombination may resolve the current $2\sigma$ tension between the BBN and CMB determinations of the baryon energy density.
Future CMB-S4 measurements will be able to probe this scenario directly.
\end{abstract}
\maketitle

\section{Introduction}

Recently, the Experiment to Detect the Global Epoch of Reionization Signature (EDGES)~\cite{Bowman:2018yin}, which targets the global 21cm signal, announced a detection of an absorption profile centered at $78\,{\rm MHz}$ (corresponding to redshift $z\!\sim\!17$ for 21cm line emission from neutral hydrogen), with a best-fit amplitude more than twice the maximum allowed in the standard cosmological model ($\Lambda$CDM). 
As predicted in Refs.~\cite{Tashiro:2014tsa,Munoz:2015bca}, elastic scattering between dark matter (DM) and baryons with a Rutherford cross section ($\propto v^{-4}$, where $v$ is the relative velocity between the particles) can result in considerable cooling or heating of the baryon gas, potentially altering the expected 21cm signal at high redshift. Concurrent with the EDGES announcement, Ref.~\cite{Barkana:2018lgd} attributed the large absorption amplitude to gas temperatures that were significantly below the $\Lambda$CDM prediction, and suggested that the gas cooling resulted from a DM--baryon Coulomb-type interaction. 

In the context of full particle-physics models, the phenomenological interaction with a $v^{-4}$ cross section corresponds to DM interacting with baryons through a mediator that is massless or has a mass much smaller than the momentum transfer. Several studies have explored the DM-scattering interpretation of the EDGES signal and have identified a viable parameter space for millicharged DM comprising a small fraction of the total DM to explain it~\cite{Munoz:2018pzp,Berlin:2018sjs,Barkana:2018qrx}. Meanwhile, severe constraints were placed on other potential models involving a light mediator~\cite{Barkana:2018qrx}.

The appeal of $v^{n}$ interactions with $n\!=\!-4$ is that their effect is more important at late times compared to higher values of $n$. In fact, they are most effective during the Cosmic Dawn era---which the EDGES signal corresponds to---as this marks the lowest point throughout the history of the Universe for the globally-averaged baryon temperature. Nevertheless, such interactions could certainly leave a detectable imprint in the small-scale temperature and polarization fluctuations of the cosmic microwave background (CMB) radiation~\cite{Dvorkin:2013cea}. 
In Ref.~\cite{paper1}, we present new limits from {\it Planck} 2015 data on late-time DM--proton interactions, extending former studies~\cite{Xu:2018efh,Slatyer:2018aqg} to the case of a strongly-coupled sub-component of the DM. 

In this work we perform a thorough investigation of the 21cm phenomenology of DM--baryon interactions during Cosmic Dawn. We demonstrate that, 
depending on the interaction cross section, there are three distinct regimes: (i) for weak coupling, the baryons cool adiabatically (due to the Hubble expansion), with at most a slight enhancement due to energy continuously lost to heating the DM; (ii) for a narrow range of stronger coupling strengths, the cooling effect reaches peak efficiency. In this regime the baryons and the DM reach equilibrium before or during Cosmic Dawn, but the DM is still significantly cooler than the baryons prior to their decoupling from the CMB; (iii) for strong coupling, DM is tightly coupled to the baryons (and thus behaves as an additional component of the baryon fluid, as described in Ref.\cite{paper1}) prior to recombination, 
thereby shutting off the subsequent non-adiabatic cooling mechanism of the gas. As we will show, the 21cm absorption minimum then saturates for higher cross sections, at a level which depends on the interacting DM fraction and its mass.
As a result of this behavior, there exists for each mass a lower limit on the fraction below which the gas cooling cannot explain the EDGES signal. 

Meanwhile, based on the  CMB limits found  in Ref.~\cite{paper1}, we show  that---augmenting previous claims~\cite{Munoz:2018pzp,Barkana:2018qrx,Berlin:2018sjs}---not even a percent of the total DM can be millicharged at the level needed to explain EDGES. It is only for fractions $f_{\chi}$ of interacting DM less than $f_{\chi}\simeq0.4\%$  (a value slightly lower than Ref.~\cite{dePutter:2018xte}) that a window opens up for substantial DM--baryon interactions, as any CMB experiment is not sensitive to values less than (or very close to) its fractional uncertainty in the baryon energy density. 

In light of the lower and upper limits on the fraction of interacting DM, and given additional constraints on DM millicharge, most notably from stellar cooling~\cite{Vogel:2013raa}, cooling of supernova 1987A~\cite{Chang:2018rso}, and a search for millicharged particles in SLAC~\cite{Prinz:1998ua}, we derive the tightest limits on this model to date.
We find that only a minimal region in parameter space remains consistent with the EDGES result, namely a fraction of millicharged DM between $0.0115\%\left(m_\chi/1\,{\rm MeV}\right)\lesssim f_{\chi}\lesssim 0.4\%$ with masses $m_\chi$ in the range $0.5\,{\rm MeV}-35\,{\rm MeV}$  and charge in a narrow band---whose width depends on the DM fraction and mass---within the range $10^{-6} e\lesssim\epsilon e\lesssim10^{-4} e$.  Curiously, a $f_\chi\!\sim\!0.4\%$ DM component tightly coupled to baryons at recombination could explain the excess of baryon energy density as inferred from the CMB \cite{Ade:2015xua,Aghanim:2018eyx}, compared with the Big-Bang nucleosynthesis (BBN) estimate (which is based on  deuterium abundance measurements \cite{Cooke:2017cwo,Zavarygin:2018dbk}).

Our results also confirm that in the case where the interaction is effectively modeled as $v^{-4}$ scattering with hydrogen, current CMB limits in principle do not rule out the phenomenological DM interpretation of the EDGES anomaly~\cite{Munoz:2015bca,Barkana:2018lgd}. 

However, as the viable ranges of cross sections for both the millicharged DM model and the phenomenological Coulomb-type interaction are limited---and will be further constrained by future experiments such as CMB-S4~\cite{Abazajian:2016yjj}, as we discuss below---it is important to consider the pertaining astrophysical uncertainties when evaluating the prospects to {\it realistically} explain the EDGES signal. Our discussion of these uncertainties illustrates the key role of the Lyman-$\alpha$ coupling of the 21cm spin temperature to the baryon gas temperature, and the potentially significant effect of various sources of heating. 
 
This paper is structured as follows.
In Section~\ref{sec:21cmDMSignal}, we review the equations governing the evolution of the DM and baryon temperatures, as well as their relative velocity, in the presence of $v^{-4}$ interaction. We then present the details of our calculation of the sky-averaged 21cm brightness temperature.
In Section~\ref{sec:DMInterpretation}, we reassess the implications of CMB limits, as well as other constraints, for the fractional millicharged DM interpretation of the EDGES signal, and for the fiducial direct DM-hydrogen interaction. We also present forecasts for the sensitivity of the next-generation CMB-S4 experiment to improve the limits on such interactions~\cite{paper1}.
In Section~\ref{sec:Astrophysics}, we make contact with the (astro)physical world and discuss the various uncertainties involved in making realistic estimates for the 21cm signal. We conclude in Section~\ref{sec:conclusions}.

\section{Calculation of the 21cm Global Signal with DM-Baryon Interaction}
\label{sec:21cmDMSignal}

\subsection{Late-Time DM-Baryon Scattering}
The equations for the evolution of the DM-baryon relative velocity ${V}_{\chi b}$ and the evolution of their temperatures $T_\chi$ and $T_b$ were first derived in Ref.~\cite{Munoz:2015bca}. Here we present their generalized form, to account for separate interactions with free electrons and protons when needed, and for a fraction $f_{\chi}$ of interacting DM (see \cite{Munoz:2018pzp,Liu:2018uzy, paper1}). The equation for the relative bulk velocity is given by
\begin{equation}
  \dot{V}_{\chi b} 
  = - \frac{\dot{a}}{a} V_{\chi b} - \left(1+\frac{f_{\chi}\rho_{\rm DM}}{\rho_b}\right) \sum\limits_t\frac{\rho_t\sigma_{0,t}F(r_t)}{(m_\chi + m_t)V_{\chi b}^2}\ ,
  \label{eq:V}
\end{equation}
where the dot stands for derivative with respect to proper time, $t$ stands for the target particle, $\sigma_{0,t}$ is the cross section, $\rho_{\rm DM}$ is the density of all of DM, and we  define 
\begin{align}
&r_t \equiv V_{\chi b} / u_{\chi t}, ~~~~~~u_{\chi t}\equiv\sqrt{T_\chi/m_\chi+T_b/m_t} \nonumber \\ 
&F(r_t)\equiv\left[\mathrm{Erf}\left(\frac{r_t}{\sqrt{2}}\right) - \sqrt{\frac{2}{\pi}} r_t e^{-r_t^2/2} \right]. \nonumber
\end{align}
The temperature evolution equations are given by
\begin{eqnarray}
  \dot{T}_\chi 
  &=& -2\frac{\dot{a}}{a} T_\chi +\sum\limits_t\frac{2}{3}\frac{m_\chi \rho_t \sigma_{0,t}}{u_{\chi t}^{3}(m_\chi + m_t)^2}  \nonumber \\
  &\times&\left\{\sqrt{\frac{2}{\pi}} (T_b - T_\chi) e^{-r_t^2/2} +   m_t \frac{V_{\chi b}^2}{r_{t}^3}F(r_t) \right\} \label{eq:Tchi}\\
  \dot{T}_b 
  &=& - 2\frac{\dot{a}}{a} T_b + \sum\limits_t\frac{2}{3}\frac{f_{\chi} \rho_t \rho_{\rm DM} \sigma_{0,t}}{u_{\chi t}^{3}(1+f_{H_e}+x_e)n_H(m_\chi + m_t)^2}  \nonumber \\
  &\times&\left\{\sqrt{\frac{2}{\pi}} (T_\chi - T_b) e^{-r_t^2/2}  + m_\chi  \frac{V_{\chi b}^2}{r_{t}^3} F(r_t) \right\}  \nonumber \\ 
  &+&\Gamma_C(T_{\rm CMB}-T_b), \label{eq:Tb}
\end{eqnarray}
where $T_{\rm CMB}$ is the CMB temperature, and $\Gamma_C$ is the Compton scattering rate, which depends on $x_e\equiv n_e/n_H$ and $f_{\rm He}\!\equiv\!n_{\rm He}/n_H$, 
the free-electron and helium fractions.
In the case of direct interaction with hydrogen, the sum is over a single target for which we replace $m_t$ by $m_H$, the hydrogen mass.
Lastly, we need to include the evolution equation of the free electron fraction $x_e$, which directly depends on the baryon temperature $T_b$,
\begin{eqnarray}
\frac{dx_{e}(z)}{dz}=\frac{C}{(1+z)H(z)}\bigg[\alpha_{H} x_e^2 n_H-\beta_{H}(1-x_e)e^{-\frac{h\nu_\alpha}{k_bT_{\rm CMB}}}\bigg], \nonumber \\
\label{eq:xe}
\end{eqnarray}
where the coefficients $\alpha_H(T_b,T_{\rm CMB})$ and $\beta_H(T_{\rm CMB})$ are the effective recombination and photoionization rates, $\nu_\alpha$ is the Lyman-$\alpha$ frequency, and $C$ is the Peebles factor representing the probability for an electron in the $n = 2$ state to get to the ground state before being ionized~\cite{AliHaimoud:2010dx}.  
We note that this derivation neglects the effects of structure formation, which may affect the scattering rates.\footnote{We thank Ilias Cholis for spurring a discussion on this point.}  
  
\subsection{The 21cm Brightness Temperature}

Cosmic Dawn---the birth of the first astrophysical sources---is expected to imprint a negative spectral distortion in the 21cm brightness temperature contrast with respect to the CMB, the exact details of which depend on various astrophysical processes.
This comes about as the primordial gas that permeates the Universe begins to cool adiabatically following recombination, once Compton scattering with the remaining free electrons is no longer efficient  to couple the gas temperature to that of the CMB.
As a consequence, the gas temperature starts dropping faster than that of the CMB, allowing CMB photons to be absorbed as they travel through neutral hydrogen gas clouds. The amount of absorption depends on the spin temperature of the gas, which parameterizes the ratio between the populations of the triplet and singlet states of the hyperfine transition.

The spin temperature $T_s$ during the Cosmic Dawn era is determined by competing processes which drive its coupling to either the background radiation temperature---which we take to be the CMB temperature $T_{\rm CMB}$---or to the kinetic temperature of the baryon gas, $T_b$.
These processes include collisional excitations in the gas (with hydrogen atoms, free electrons and protons), absorption of CMB photons, and photo-excitation and de-excitation of the Lyman-$\alpha$ transition by Lyman-$\alpha$ photons emitted from the first stars (which is known as the Wouthuysen-Field effect~\cite{Wouthuysen1952,Field1959}).
$T_s$ is roughly approximated by~\cite{Field1959}
\begin{equation}
T_s^{-1} \approx \frac{T_{\rm CMB}^{-1} + (x_c + x_\alpha) T_b^{-1} }{1 + (x_c + x_{\alpha})},
\label{eq:Ts}
\end{equation}
where $x_c$ is the collisional coupling coefficient~\cite{Zygelman:2010zz}, $x_{\alpha}$ the Lyman-$\alpha$ coupling coefficient~\cite{Pritchard:2011xb}, and we have set the color temperature equal to the kinetic gas temperature.

The 21cm brightness temperature contrast with respect to the CMB is given by \cite{Loeb:2003ya}
\begin{eqnarray}
T_{\rm 21}(z)&=& \frac{T_s-T_{\rm CMB}}{1+z}\left(1-e^{-\tau}\right) \nonumber \\
\tau&=&\frac{3T_{*}A_{10}\lambda_{21}^3  n_{\rm HI}}{32\pi T_s H(z)}, 
\label{eq:T21}
\end{eqnarray}
where $\tau$ is the optical depth for the transition. It depends on $n_{\rm HI}$, the neutral hydrogen density; $A_{10}$, the Einstein-A coefficient of the hyperfine
transition; $T_{*}=0.068\,{\rm K}$,  the energy difference between the two hyperfine levels; and $\lambda_{21}\approx21.1\,{\rm cm}$, the transition wavelength.
Given the large astrophysical uncertainties, there is no exact prediction for this signal during Cosmic Dawn, and different models easily generate values of $T_{\rm 21}$ in the $z=13-20$ redshift range that differ by more than an order of magnitude~\cite{Cohen:2016jbh}.
The key point to appreciate is that this quantity has a minimum value, obtained when the spin temperature equals the gas temperature. Under $\Lambda$CDM, this value is known to better than percent accuracy. Setting $T_s= T_b$ in Eq.~\eqref{eq:T21} yields $T_{\rm 21}(z\sim17)\simeq-207\,{\rm mK}$.
However, as described in Ref.~\cite{Munoz:2015bca}, this prediction can change in the presence of DM--baryon scattering, since the gas experiences a cooling effect as it deposits heat into the colder dark matter fluid, as well as a heating effect as the kinetic energy associated with the relative bulk velocity between the baryons and dark matter is dissipated following recombination. For low DM-particle masses and at Cosmic Dawn redshifts, the cooling effect is dominant.

To calculate the 21cm brightness temperature in the presence of $v^{-4}$ interactions, we follow the prescription of Ref.~\cite{Munoz:2015bca} and solve the system of equations~\eqref{eq:V}--\eqref{eq:xe}, starting the integration long before Cosmic Dawn.
For each choice of model and cross-section amplitude, we calculate $T_{\rm 21}$ in Eq.~\eqref{eq:T21} for an array of different initial relative bulk velocities $V_{{\rm \chi b},0}$ between the DM and baryons, and then calculate the observed global brightness temperature---which is an average weighted by a Maxwell-Boltzmann probability distribution function of initial relative velocities, with a root-mean-square value $V_{\rm RMS}$---according to
\begin{eqnarray}
\langle T_{\rm 21}(z)\rangle &=& \int dV_{{\rm \chi b},0}T_{\rm 21}(z,V_{{\rm \chi b},0})\mathcal{P}(V_{{\rm \chi b},0})~, \nonumber \\
\mathcal{P}(V_{{\rm \chi b},0})&=&4\pi V_{{\rm \chi b},0}^2 e^{-3V_{{\rm \chi b},0}^2/2V_{\rm RMS}^2}/(2\pi V_{\rm RMS}^2/3)^{3/2}.
\label{eq:T21V}
\end{eqnarray}
To properly take into account the baryon-photon drag at earlier times, we use initial conditions 
for Eqs.~\eqref{eq:V}-\eqref{eq:Tb} taken from the output of a full-fledged Boltzmann code (using a modified version of the \texttt{CLASS}{} package~\cite{Lesgourgues:2011re}, described in great detail in Ref.~\cite{paper1}).  As we emphasize below, in the strong-coupling regime it is imperative to track these quantities starting from well before recombination, as the $\Lambda$CDM conditions of vanishingly small $T_\chi$ and $V_{\rm RMS}=29\,{\rm km/sec}$ are no longer a valid approximation~\cite{Liu:2018uzy}. Rather, we will see that in those circumstances $T_\chi$ approaches $T_b$ (which is still coupled to $T_{\rm CMB}$) and $V_{\rm RMS}$ approaches zero (see Appendix for more detail). It is the very delicate balance between the different rates---Hubble, Compton and DM-baryon heat exchange---that governs the behavior of the 21cm Cosmic Dawn signal.

The EDGES  collaboration recently reported a best-fit minimum  temperature value with $99\%$-confidence bounds of $T_{\rm 21}(\nu\!=\!78.1\,{\rm MHz})=-500^{+200}_{-500}\,{\rm mK}$~\cite{Bowman:2018yin}, a $3.8\sigma$ deviation from the minimum value 
under $\Lambda$CDM at this observing frequency. 
Therefore, unless stated otherwise, our criterion used throughout for consistency with the EDGES measurement is that $\langle T_{\rm 21}(\nu\!=\!78.1\,{\rm MHz})\rangle\!=\!-300\,{\rm mK}$ (corresponding to the EDGES upper bound), and is calculated when setting the spin temperature equal to the velocity-dependent baryon temperature, i.e.~$T_s(z)= T_b(V_{{\rm \chi b},0}, z)$.  We set the cosmological parameters to best-fit values derived from the same {\it Planck} 2015 ``TT+TE+EE+lensing" dataset used in Ref.~\cite{paper1}. 

\section{Constraints on the DM Interpretation of the EDGES Signal}
\label{sec:DMInterpretation}

\subsection{Millicharged DM}

If DM has a millicharge under electromagnetism, its momentum-transfer cross section with free electrons or protons (denoted by $t$) is $\sigma_t=\sigma_{0,t} v^{-4}$, with
\begin{equation}
\sigma_{0,t}= \frac{2\pi\alpha^2\epsilon^2\xi}{\mu_{\chi,t}^2}, ~~~~~~~~~~~ \xi=\log{\left(\frac{9T_b^3}{4\pi\epsilon^2\alpha^3 x_e n_H}\right)},
\label{eq:DMmillicharge}
\end{equation}
where $\alpha$ is the fine-structure constant, $\mu_{\chi,t}$ is the reduced mass, and the factor $\xi$ arises from regulating the forward divergence of the differential cross section through Debye screening.
This scenario for EDGES is comparatively easier to constrain with the CMB, as millicharged DM only interacts with the (charged) ionized particles.
The effects on the CMB originate mostly from a time prior to recombination when the cosmic plasma was fully ionized~\cite{paper1, Slatyer:2018aqg}.
However, by the onset of Cosmic Dawn, the ionization fraction of the gas is very small ($\sim2\times10^{-4}$), suppressing the efficiency of the DM--baryon interaction in cooling the baryon gas. As a result, the required cross section to explain EDGES with $100\%$ millicharged DM is orders of magnitude larger than the CMB limit\footnote{Note that in calculating the CMB limits in Ref.~\cite{paper1}, we neglected interactions with free electrons and helium. As these would only strengthen the constraints, our conclusions are conservative.}. Next, we will first compare the two for $f_{\chi}\!=\!1\%$, and then proceed to investigate lower fractions---in particular lower than the fractional uncertainty on the baryon energy density---which are poorly constrained by the CMB.

\vspace{-0.1in}
\subsubsection*{$f_{\chi}\sim1\%$ millicharged DM}
\begin{figure}
\centering
\includegraphics[width=\columnwidth]{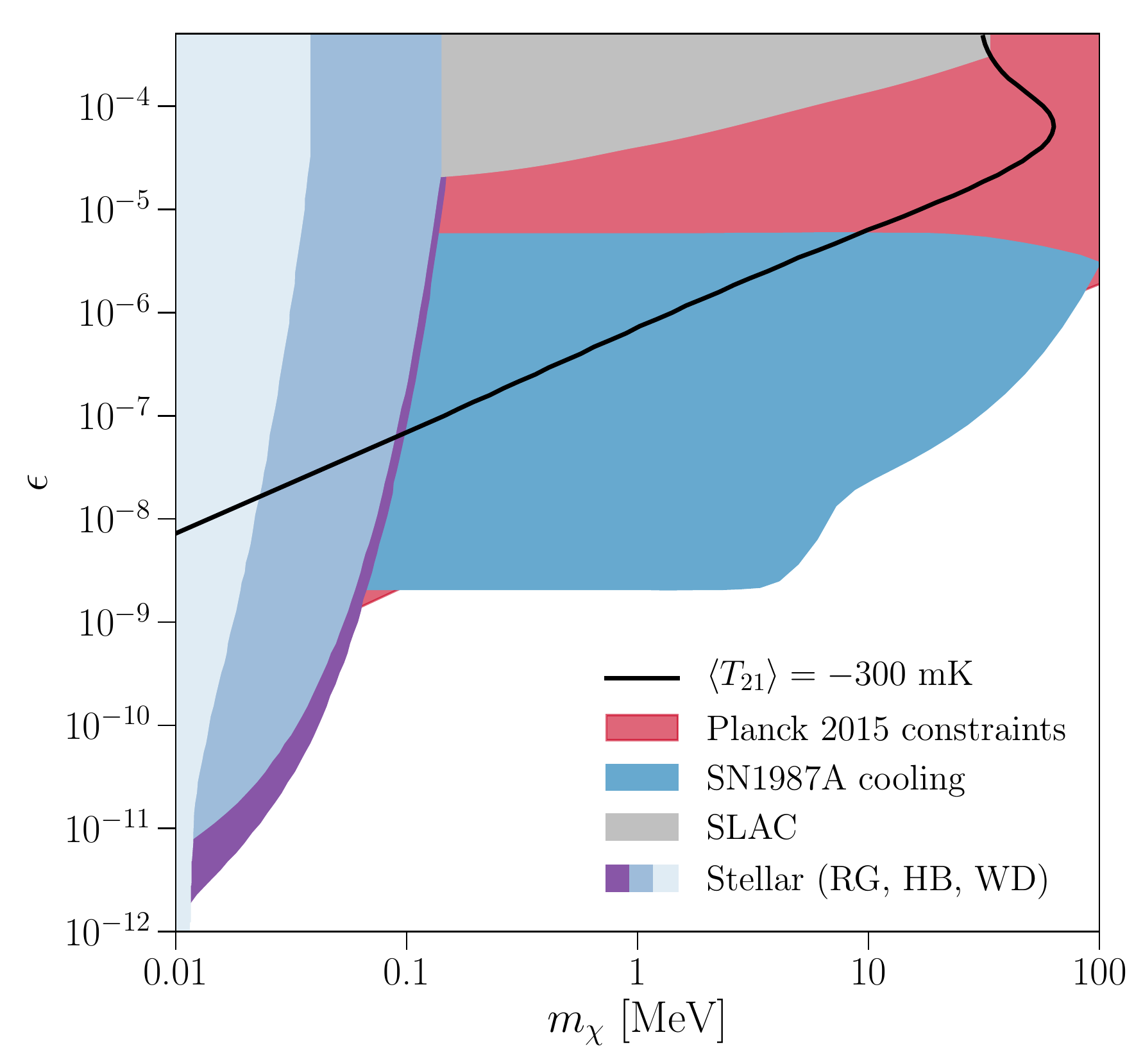}
\caption{$f_{\chi}=1\%$ millicharged DM. We show the allowed region for the charge fraction $\epsilon$ to explain the EDGES signal ({\it black}) as well as limits from cooling of SN1987A~\cite{Chang:2018rso} ({\it blue}), the SLAC millicharge experiment~\cite{Prinz:1998ua} ({\it gray}), {\it Planck} 2015 CMB data~\cite{paper1} ({\it pink}) and stellar (Red Giant, Helium Burning and White Dwarf) cooling~\cite{Vogel:2013raa} ({\it various shades}). As explained in the text, for masses above $\sim85\,{\rm MeV}$, the minimum 21cm brightness temperature never falls below $\langle T_{21}\rangle=-300\,{\rm mK}$.}
\label{fig:onepercentConstraints}
\end{figure}

\vspace{-0.025in}
As shown in Figure~\ref{fig:onepercentConstraints}---and in agreement with Ref.~\cite{Munoz:2018pzp}---interaction with $f_{\chi}\!=\!1\%$ millicharged DM with a charge fraction larger than $\epsilon\simeq6.2\times10^{-7}\left(m_\chi/{\rm Mev}\right)$ could in principle cool the baryon gas temperature enough to explain the EDGES 21cm measurement, as long as the DM-particle mass is lower than $\sim85\,{\rm MeV}$. For higher masses, the required cross sections are in the strong coupling regime, where the cooling efficiency hits a ceiling (more on this below).  
As shown in Figure~\ref{fig:onepercentConstraints}, the $95\%$-C.L.\ CMB upper limits from Ref.~\cite{paper1}, $\epsilon\lesssim1.8\times10^{-8}(\mu_{\chi,p}/{\rm MeV})^{1/2}(m_\chi/{\rm MeV})^{1/2}$, are more than an order of magnitude lower.\footnote{To convert the limit on the cross-section derived in Ref.~\cite{paper1} to a limit on the charge fraction $\epsilon$, we use Eq.~(\ref{eq:DMmillicharge}) at $z=1100$. We note that this is not strictly exact since this relation is temperature dependent and the CMB probes a relatively wide redshift range. However, it leads to a conservative upper-limit on $\epsilon$ since the cross section drops logarithmically as $T_b$ decreases with time.} The corresponding cross sections are two orders-of-magnitude discrepant. Therefore $f_{\chi}=1\%$ millicharge DM is strongly ruled out.

\vspace{-0.2in}
\subsubsection*{Sub-percent fractions of strongly-coupled millicharged DM}
\vspace{-0.025in}

Given the prohibitive CMB constraints on $f\gtrsim1\%$ millicharged DM, we are forced to consider lower DM fractions. 
This has to be done carefully, though, as there are crucial subtleties that come into play. 
The first has to do with the CMB limits. Intuitively, it is clear that if the DM component that we surmise behaves effectively like baryons, yet has a fractional abundance that is lower than the fractional uncertainty on the baryon energy density, the CMB will not be sensitive to its presence. As we demonstrated in Ref.~\cite{paper1}, below $f_{\chi}\sim0.4\%$, the effect on the CMB power spectrum is undetectable by {\it Planck}.

\begin{figure*}
\centering
\includegraphics[width=0.32\linewidth]{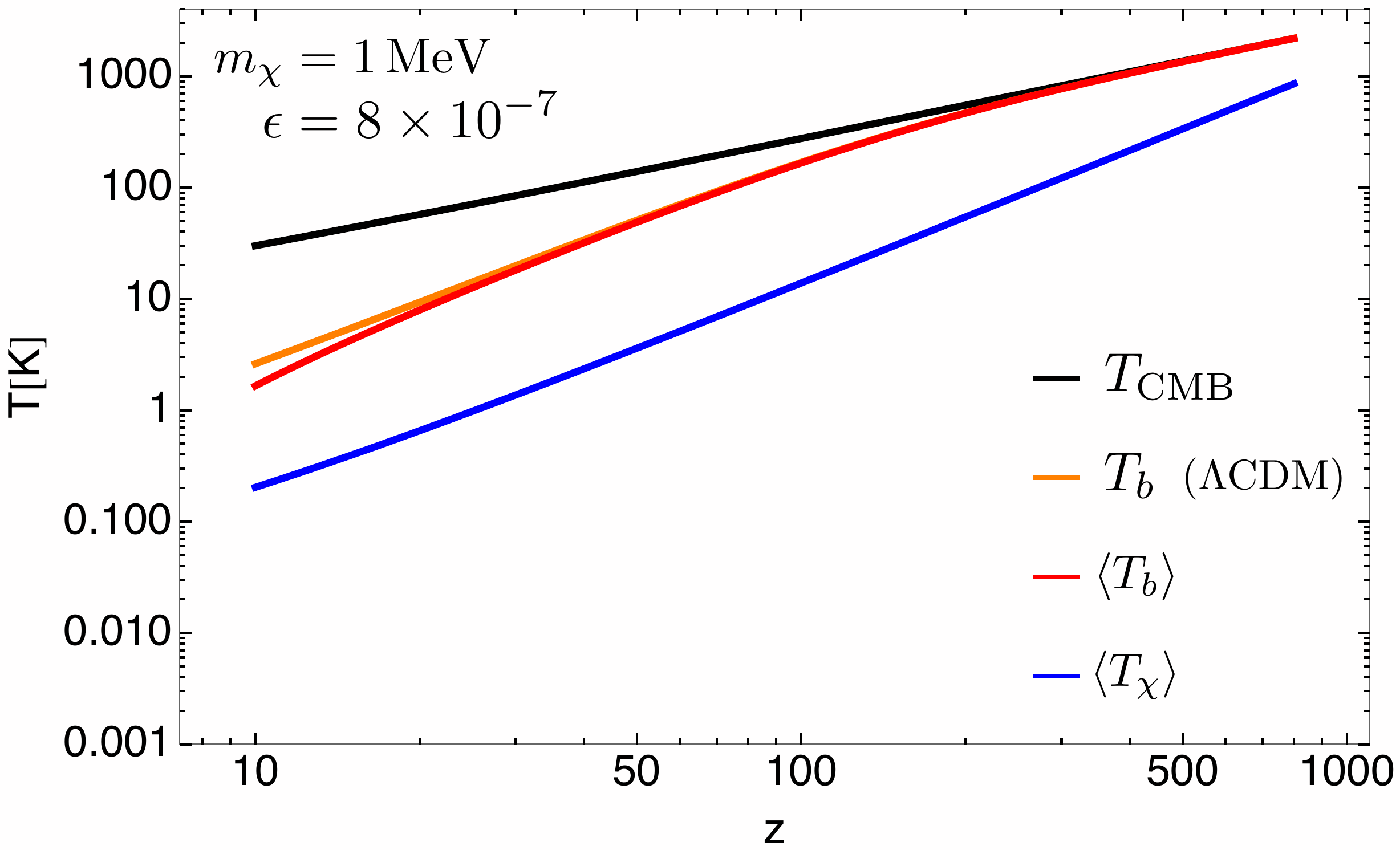}
\includegraphics[width=0.32\linewidth]{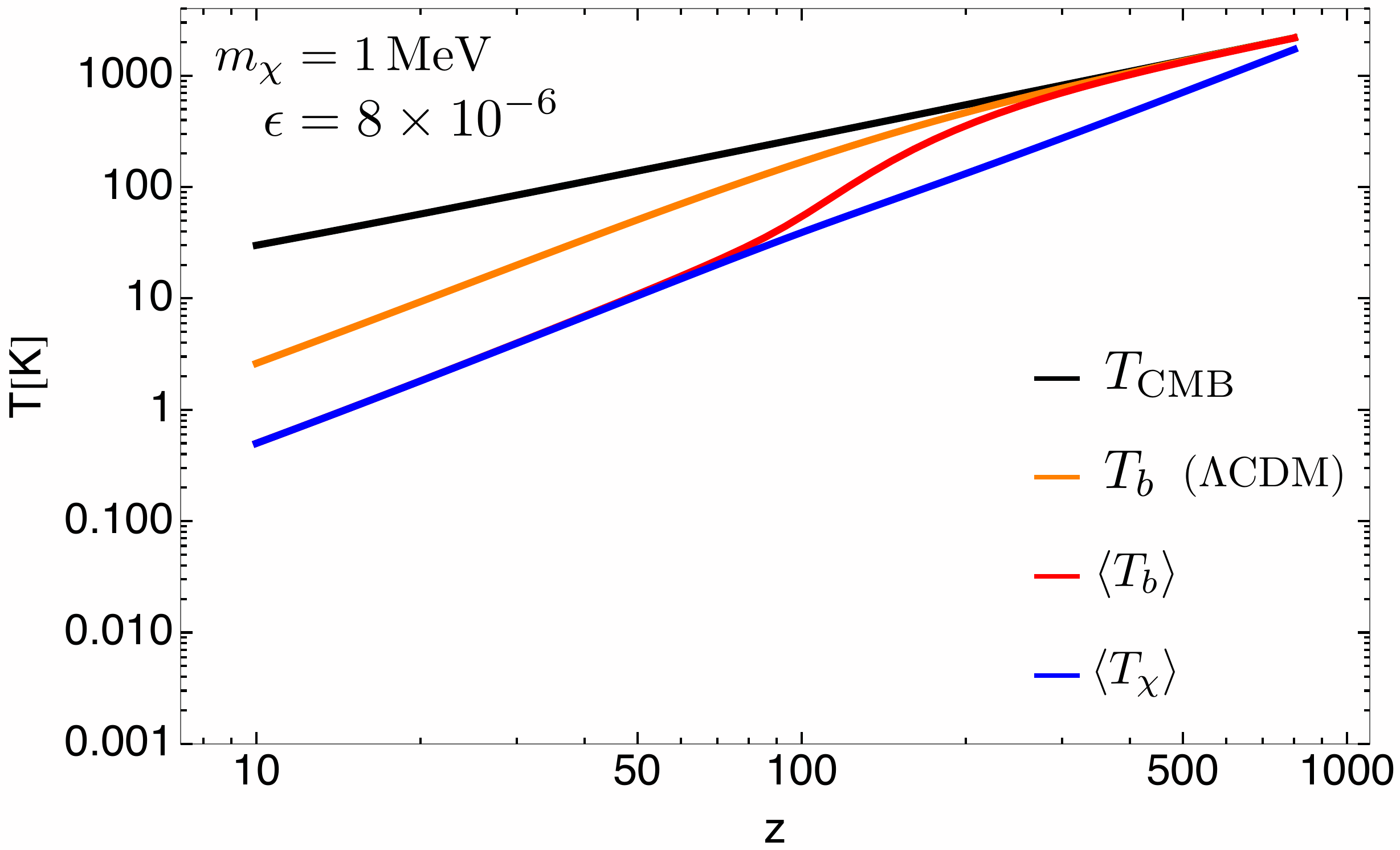}
\includegraphics[width=0.32\linewidth]{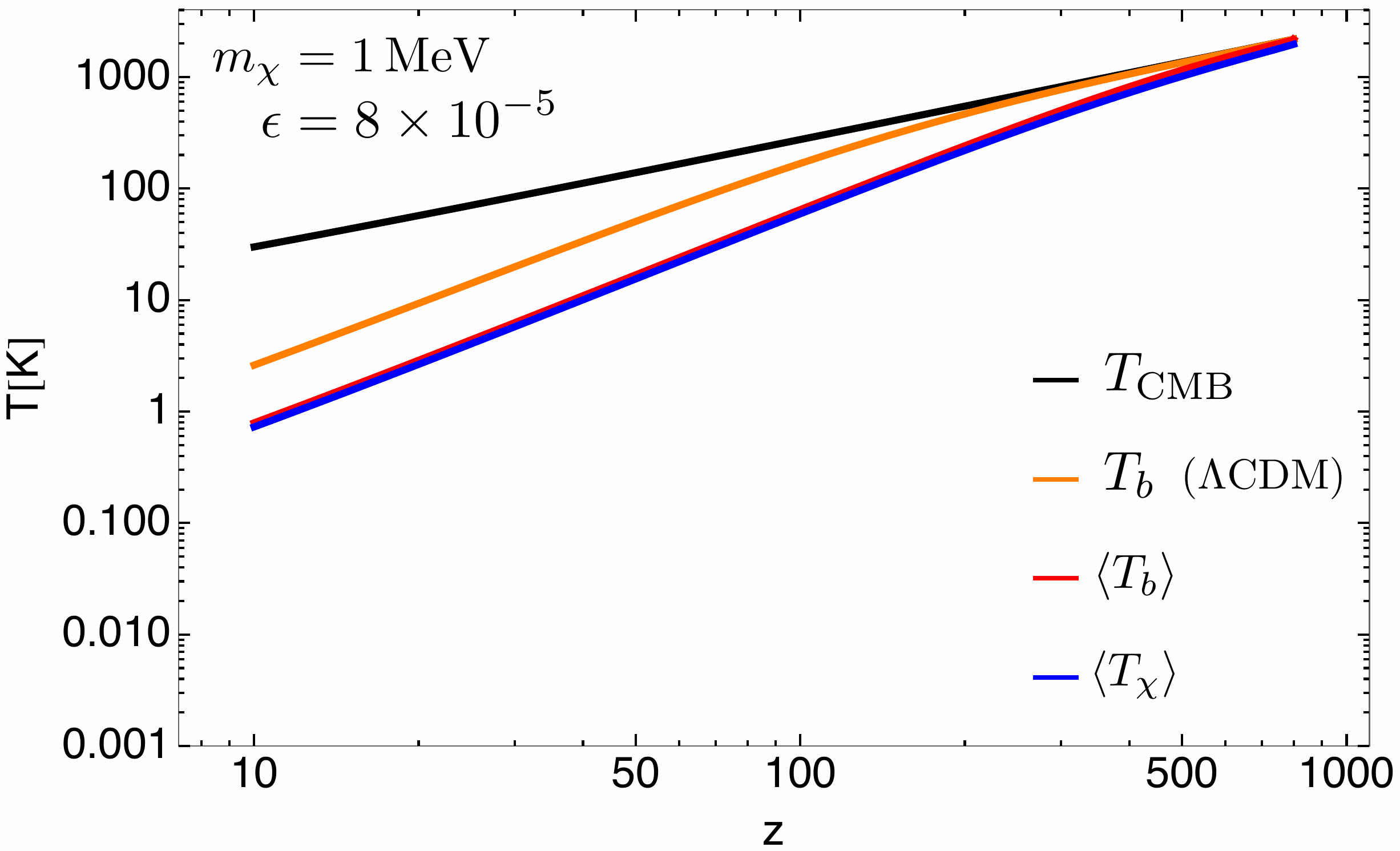}
\par\vspace{-0.1in}
\caption{The temperatures of the CMB, baryons, and DM, as a function of redshift, for different cross sections parametrized by $\epsilon$. We use a mass $m_\chi=1\,{\rm MeV}$ and a fraction $f_\chi\!=\!0.4\%$, corresponding to the black curve in the middle panel of Fig.~\ref{fig:21cmAbsTFraction}. For weak coupling, the cooling effect is limited. As the coupling increases, the effect dramatically rises, but then as the strong coupling limit is approached, the DM initial temperature reaches that of the baryons (and CMB) and the gas cooling saturates.}
\label{fig:Tregimes}
\end{figure*}

The second issue is that for decreasing DM fractions, increasingly stronger cross sections are needed in order to effectively cool the baryons. The 21cm signal is governed by the balance between the different heat exchange rates between the fluids \cite{Munoz:2015bca,Liu:2018uzy}. For the DM, as evident from Eq.~\eqref{eq:Tchi}, the competition is between the Hubble rate and heating by baryons. For the baryons, the amount of cooling depends on when they decouple from the CMB. The decoupling redshift in turn depends on the balance between the Compton rate $\Gamma_C$ and the cooling rate, which is proportional to the cross-section amplitude $\sigma_0$ {\it and}  the fraction of interacting DM $f_{\chi}$, see Eq.~\eqref{eq:Tb}.

Generically, there are three distinct regimes for the effect of DM-baryon interactions on the 21cm absorption signal at Cosmic Dawn, as described in the introduction. These are shown in Figure~\ref{fig:Tregimes}, for a DM fraction $f_{\chi}\!=\!0.4\%$ and mass $1\,{\rm MeV}$.
For low values of the charge fraction $\epsilon$, there is at most a moderate cooling of the baryons. For a limited range of higher charges, the effect reaches a maximum. It then turns over and saturates when the DM and baryon temperatures are already tightly coupled before the latter decouples from the CMB temperature.

The implication of this behavior is that for each mass there exists a lower limit on the fraction of interacting DM below which the scattering with DM no longer leads to a deviation from the $\Lambda$CDM prediction. We find that this limit is linear in the DM-particle mass, and is approximately given by $f_{\chi}\gtrsim0.0115\%\left(m_\chi/{\rm MeV}\right)$.

\begin{figure}
\centering
\includegraphics[width=\columnwidth]{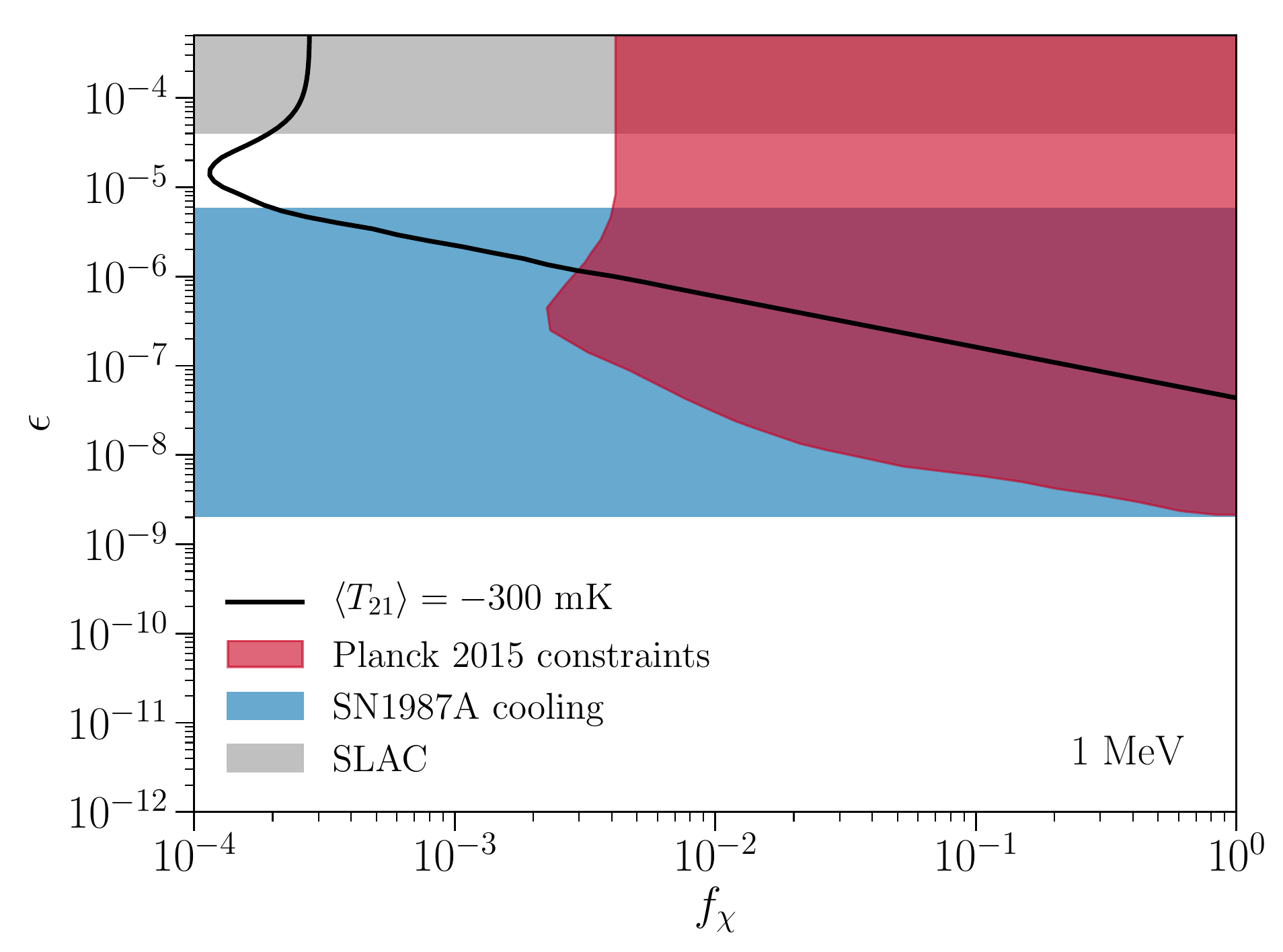}
\includegraphics[width=\columnwidth]{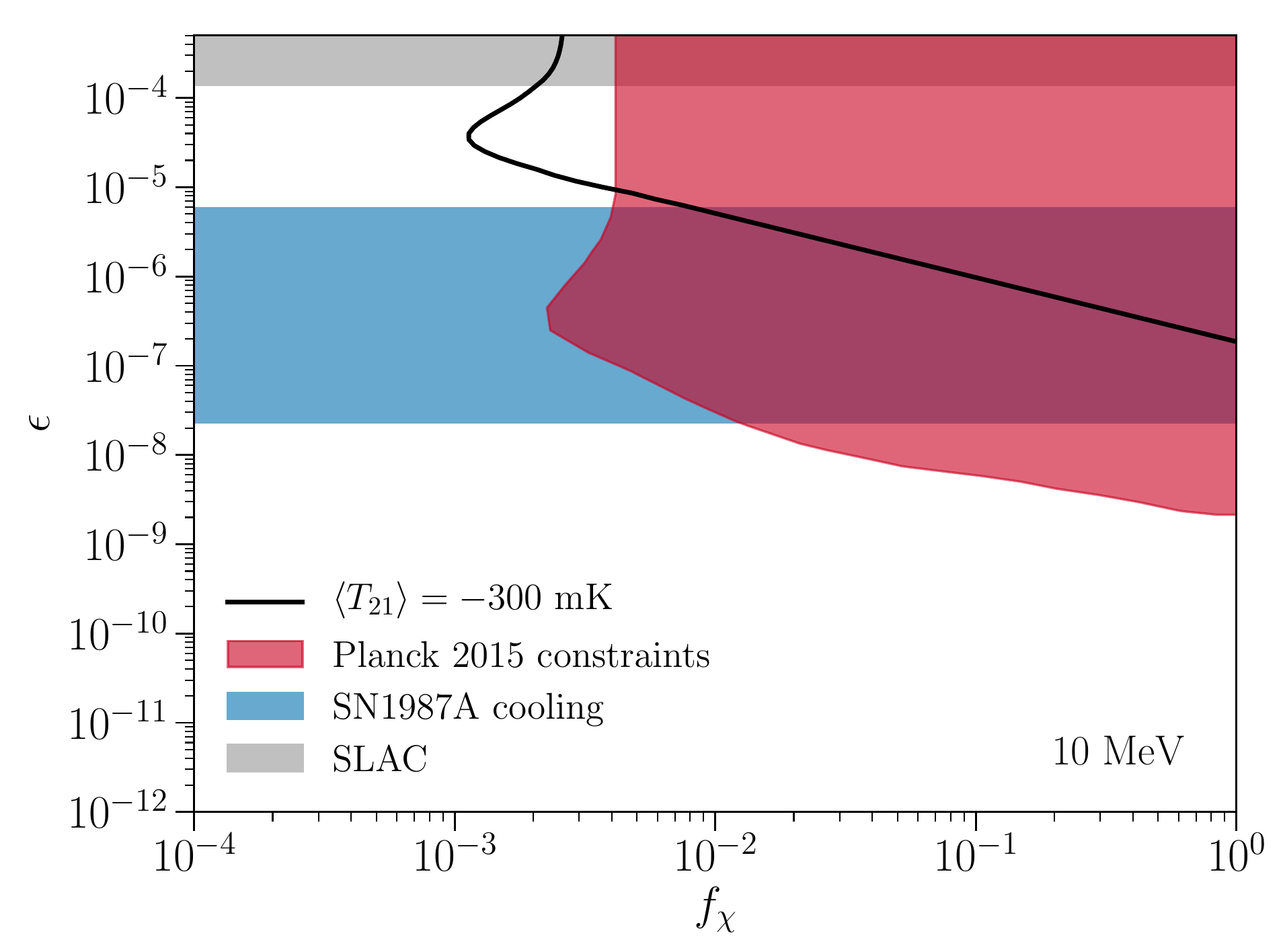}
\caption{Constraints on the charge fraction $\epsilon$ for different DM-particle masses, as a function of the DM fraction $f_{\chi}$. In comparison, we show the allowed region of charge fractions to yield $\langle T_{\rm 21}(\nu\!=\!78.1\,{\rm MHz}) \rangle\!=\!-300\,{\rm mK}$. For these masses, a window is left open between $10^{-6}\lesssim\epsilon\lesssim10^{-4}$ in the range $0.0115\%\left(m_\chi/{\rm MeV}\right)\lesssim f_{\chi}\lesssim0.4\%$. More details in Figs.~\ref{fig:21cmAbsTFraction},~\ref{fig:DMMillichargeParamSpace}.}
\label{fig:21cmConstraints}
\end{figure}

In Figure~\ref{fig:21cmConstraints}, we plot the range of charge fractions which is consistent with the EDGES result, as a function of the interacting-DM fraction, for three different masses in the range allowed by the stellar cooling bounds. We also show the limits from SN1987A cooling~\cite{Chang:2018rso} and from a search for millicharged particles in SLAC~\cite{Prinz:1998ua}, which are independent of the cosmic abundance of interacting DM (they restrict particle production, not abundance). 

We see that in all cases, the lower limit on the charge lies almost entirely within the regions ruled out by either the CMB or SN1987A.
However, the gap between the SN1987 and SLAC limits for fractions below $f_{\chi}\!\sim\!0.4\%$ potentially allows for an explanation of the EDGES anomaly as the charge is cranked up. If the transition to the strong coupling regime---where the baryon cooling effect saturates---occurs before reaching that gap, the EDGES result can be explained if the resulting 21cm absorption plateau is consistent with its uncertainty range. 

\begin{figure}
\centering
\par\vspace{0.1in}
\includegraphics[width=\columnwidth]{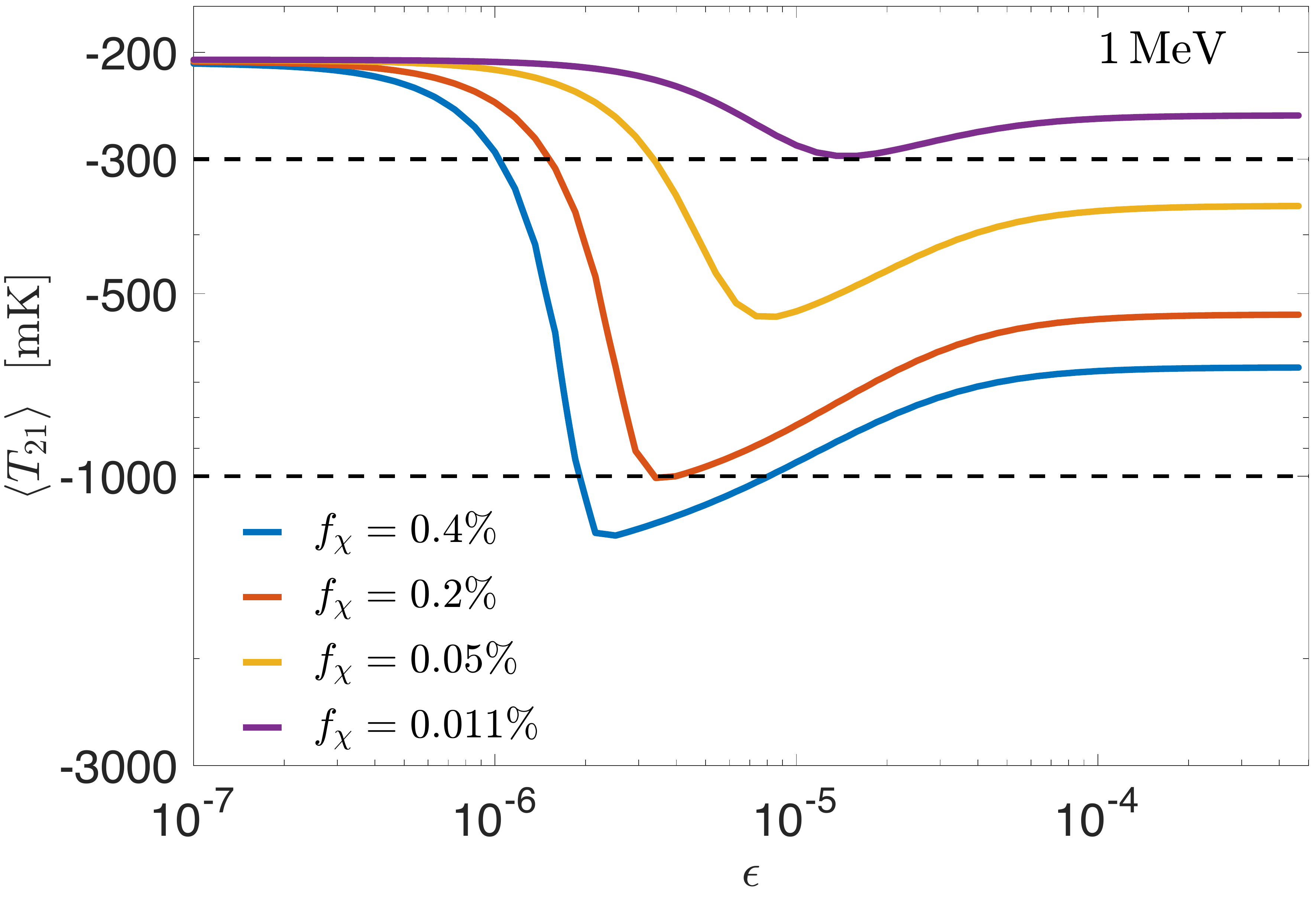}
\par
\vspace{0.2in}\includegraphics[width=\columnwidth]{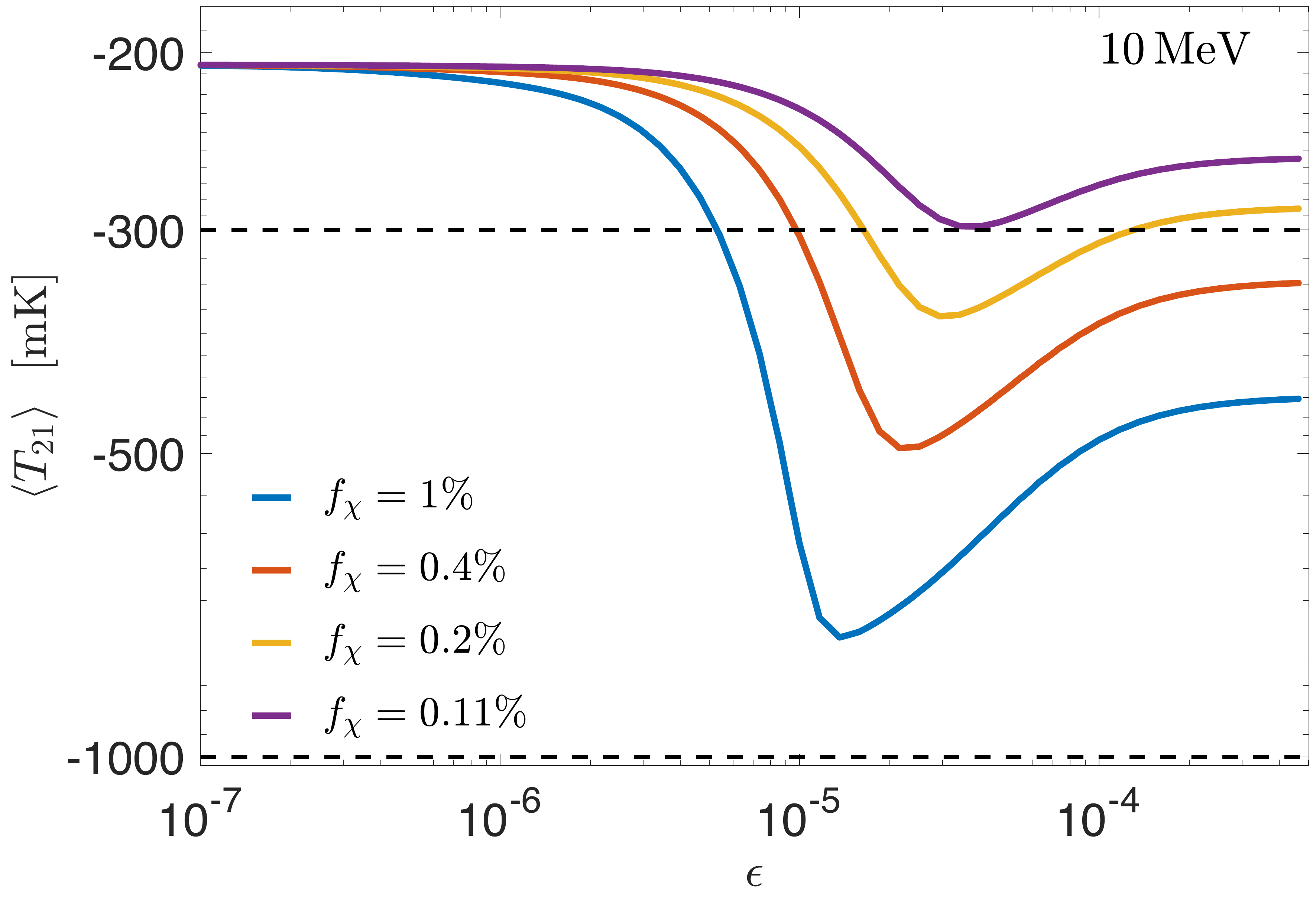}
\par
\vspace{0.075in}
\caption{The 21cm brightness temperature as a function of the charge fraction $\epsilon$, for different interacting DM fractions  $f_{\chi}$. All curves have a turnover point due to the transition into the strong-coupling regime. For $f_{\chi}\!=\!0.0115\%\left(m_\chi/{\rm MeV}\right)$, the peak absorption barely crosses the EDGES $99\%$ upper bound $\langle T_{\rm 21}(\nu\!=\!78.1\,{\rm MHz}) \rangle\!=\!-300\,{\rm mK}$. Compare against Figs.~\ref{fig:21cmConstraints}, \ref{fig:DMMillichargeParamSpace}. }
\label{fig:21cmAbsTFraction}
\end{figure}

\begin{figure}[h!]
\centering
\includegraphics[width=\columnwidth]{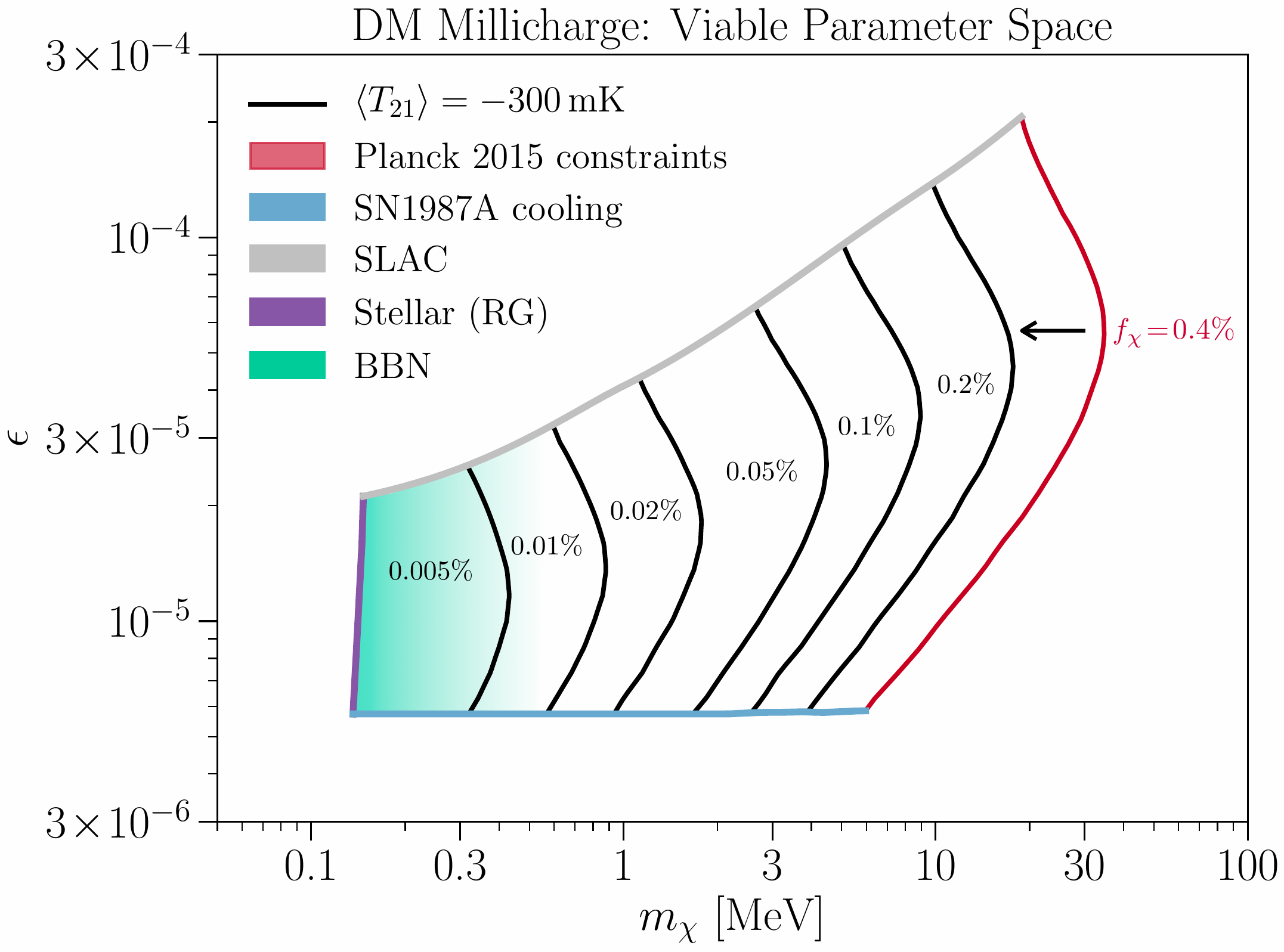}
\par\vspace{0.1in}
\includegraphics[width=\columnwidth]{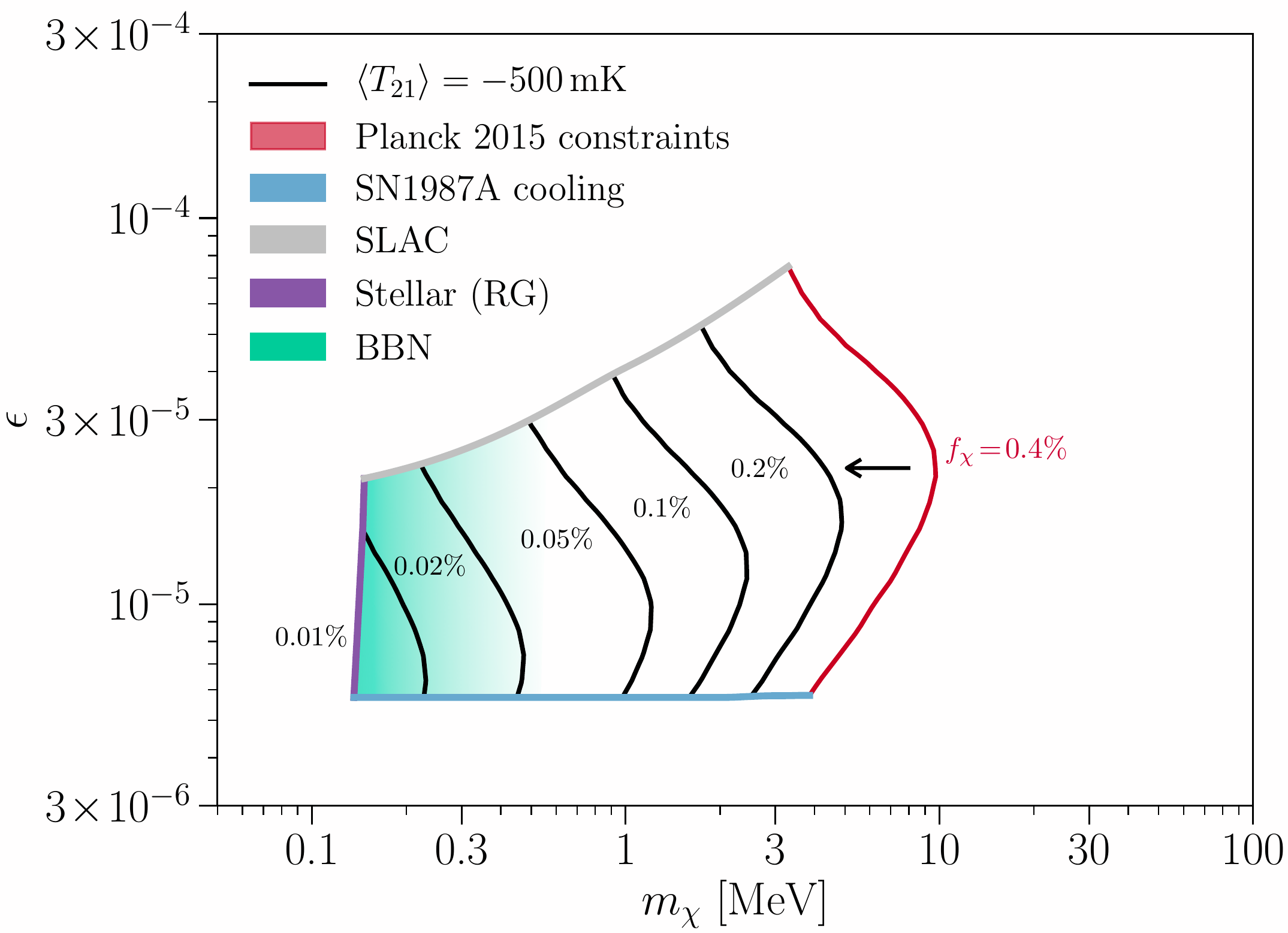}
\caption{The viable parameter space for millicharged DM to explain the anomalous EDGES 21cm signal. 
The allowed region is bound from above by SLAC constraints~\cite{Prinz:1998ua} ({\it gray}), from the left by stellar cooling~\cite{Vogel:2013raa} ({\it purple}), from below by SN1987A cooling~\cite{Chang:2018rso} ({\it blue}), and from the right by the requirement to cool the baryons enough to yield a 21cm brightness temperature consistent with the EDGES $99\%$ upper bound, $\left|\langle T_{\rm 21}(\nu\!=\!78.1\,{\rm MHz}) \rangle\right|\!=\!300 \,{\rm mK}$ \cite{Bowman:2018yin} ({\it black}). Contours are shown for several values of the fraction $f_\chi$ of the total DM that is millicharged; each yields an upper bound on the mass $m_\chi \simeq \left(f_\chi/0.0115\%\right)\,{\rm MeV}$. The rightmost limit is from Planck 2015 \cite{paper1} ({\it red}). A portion ruled out by the $N_{\rm eff}$ limit at BBN~\cite{Davidson:2000hf}, valid below $m_\chi\!\sim \!m_e$, is sketched ({\it light green}).}
\label{fig:DMMillichargeParamSpace}
\end{figure}

To explore this in more detail, we plot in Figure~\ref{fig:21cmAbsTFraction} the observed 21cm brightness temperature at $\nu\!=\!78.1\,{\rm MHz}$ as a function of the charge fraction $\epsilon$, for interacting-DM fractions in the allowed range $0.0115\%\left(m_\chi/{\rm MeV}\right)\!\lesssim\! f_{\chi}\!\lesssim\!0.4\%$. The horizontal lines indicate the $99\%$ EDGES uncertainty band, ranging from $-1000\,{\rm mK}$ to $-300\,{\rm mK}$.  
In Figure~\ref{fig:DMMillichargeParamSpace} we show the allowed region in the $(\epsilon,m_\chi)$ parameter space for several choices of the DM fraction $f_\chi$ (all below the CMB limit, i.e.~$<0.4\%$). This region is determined by the constraints plotted in Figure~\ref{fig:onepercentConstraints} (which are due to production of DM particles and are therefore independent of the fraction $f_\chi$), as well as the requirement to achieve consistency with EDGES in each case.

We note that a lower bound on the mass, $m_\chi\!\gtrsim\!10\,{\rm MeV}$, from BBN and {\it Planck} constraints on the effective number $N_{\rm eff}$ of relativistic degrees of freedom~\cite{Boehm:2013jpa}, would exclude some of this remaining region of parameter space for fractional millicharged DM~\cite{Munoz:2018pzp,Barkana:2018qrx,Berlin:2018sjs}. This bound only holds if this component was in full thermal equilibrium with electrons and photons at BBN. The $v^{-4}$ interaction we focus on here, however, is not active at early times (unless the cross sections are very large), so this would require another mechanism to couple the DM and baryon fluids. Ref.~\cite{Davidson:2000hf} derived quite stringent bounds on milicharged DM, $\epsilon<2.1\times10^{-9}$, from the $N_{\rm eff}$ constraint during BBN without assuming full equilibrium (which is the relevant scenario in the context of freeze-in DM models). However, this bound is valid in the regime $m_\chi\!\lesssim\!m_e$ (as the DM particle mass was neglected). It is therefore not reliable for most of the mass range we consider here (we investigate $m_\chi\!>\!100\,{\rm keV}$, already fairly close to $m_e$). We did not show the BBN constraints in Figures~\ref{fig:onepercentConstraints},~\ref{fig:21cmConstraints}, but do indicate its effect in Figure~\ref{fig:DMMillichargeParamSpace}. Future work to extend this bound to the full relevant mass regime is encouraged. 

The conclusion from our analysis is that the viable parameter space for explaining the anomalous EDGES measurement is limited to DM millicharge fractions $0.0115\%\left(m_\chi/{\rm MeV}\right)\!\lesssim\! f_{\chi}\!\lesssim\!0.4\%$, mass $0.5\,{\rm MeV}\!-\!35\,{\rm MeV}$ and charge in a narrow band---whose width depends on the fraction and mass---within $10^{-6} e\!\lesssim\!\epsilon e\!\lesssim\!10^{-4} e$. 
 As small as this parameter space is, it still only serves as an optimistic estimate, since we have neglected the influence of astrophysical effects, discussed separately below.
  
\vspace{-0.1in}

\subsection{Direct DM interaction with the hydrogen gas}

Albeit no such concrete particle physics model exists, if DM were allowed to interact with the neutral hydrogen itself, a much weaker cross-section amplitude might suffice to explain the EDGES results. This is because in such case there is no suppression of the scattering by the small ionization fraction during Cosmic Dawn, unlike for millicharged DM. As described earlier, when calculating the brightness temperature for DM--hydrogen scattering, we assume the DM interacts with particles of mass $m_H$.

\begin{figure}
\centering
\includegraphics[width=\columnwidth]{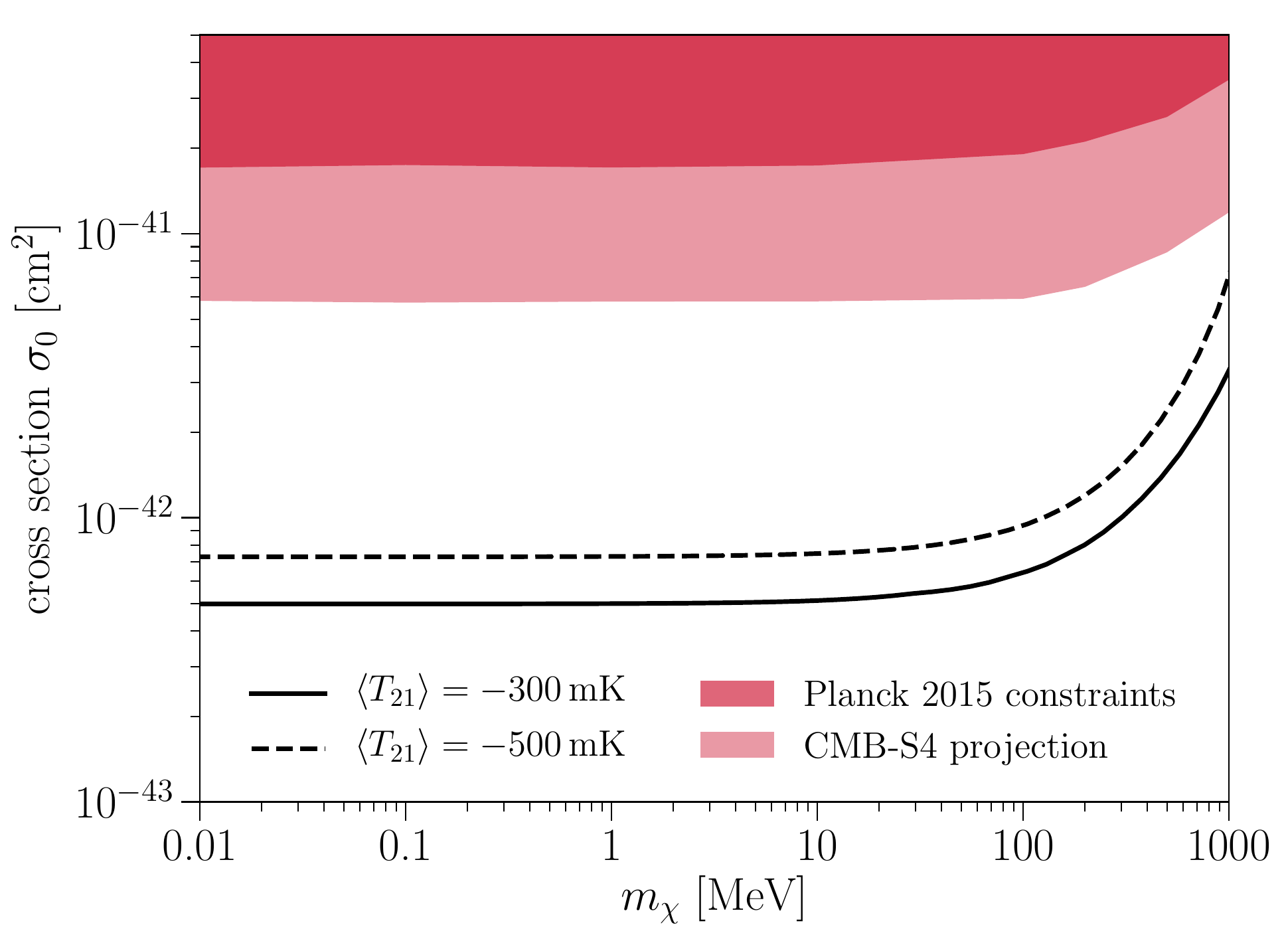}
\caption{Constraints from the CMB on the $f_{\chi}=100\%$ DM--hydrogen scenario, as a function of the DM-particle mass. We show $95\%$ C.~L.~excluded regions for DM--proton interaction, from Ref.~\cite{paper1}. The constraints ({\it in red}) were obtained from \textit{{\it Planck}} 2015 temperature, polarization, and lensing power spectra. We also show a projection of future constraints from CMB-S4~\cite{paper1} ({\it pink}), stronger by a factor $\sim3$ over most of the mass range. The lower limit on the cross section required to explain the EDGES signal ({\it black}) lies below these limits.}
\label{fig:CMBConstraints}
\end{figure}

In Figure~\ref{fig:CMBConstraints} we plot the minimal cross section---as a function of the DM particle mass---required to yield 
consistency with the EDGES $99\%$-confidence upper bound. 
The resulting curve lies  below the CMB upper limit from {\it Planck}. We also plot a forecast for CMB-S4 limits~\cite{paper1}. Evidently, there seems to be room for a phenomenological DM interpretation of EDGES even if CMB-S4 does not detect any imprint\footnote{Note that the forecast in Ref.~\cite{paper1} is conservative, as it does not include CMB lensing analysis, which may drive the constraint with future measurements~\cite{Li:2018zdm}.}. However, it is important to reiterate that for this calculation we assumed the Lyman-$\alpha$ coupling $x_\alpha$ is infinite. As this condition is purely unphysical, it is worthwhile to investigate in more detail the potential influence on our conclusions of (the  uncertain) astrophysical processes which may take effect during the Cosmic Dawn epoch. We do this in the next section.

\section{Discussion of Real-World Uncertainties}
\label{sec:Astrophysics}
\subsection{A minimal yet realistic scenario}

\begin{figure}[t]
\centering
\includegraphics[width=\columnwidth]{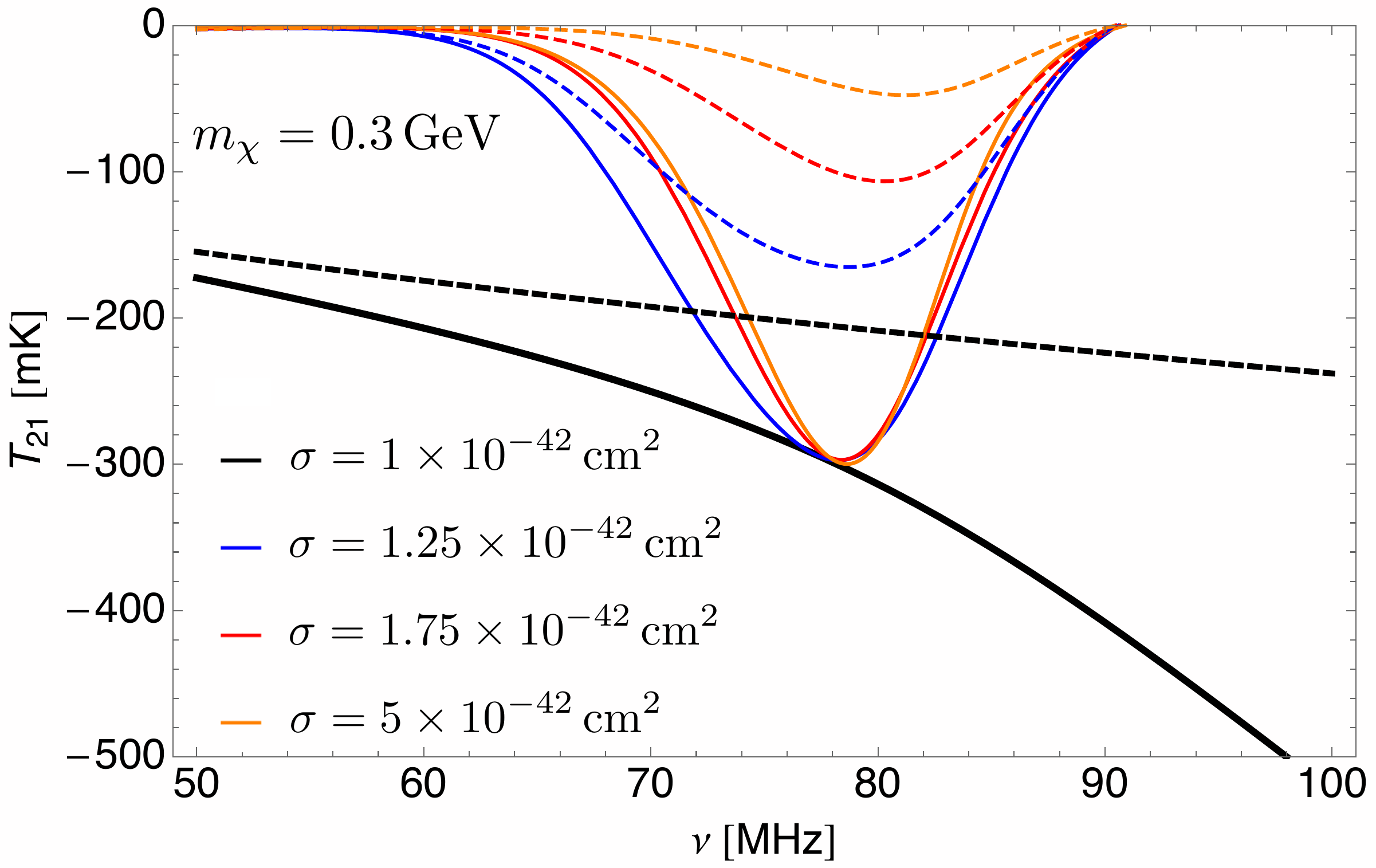}
\includegraphics[width=\columnwidth]{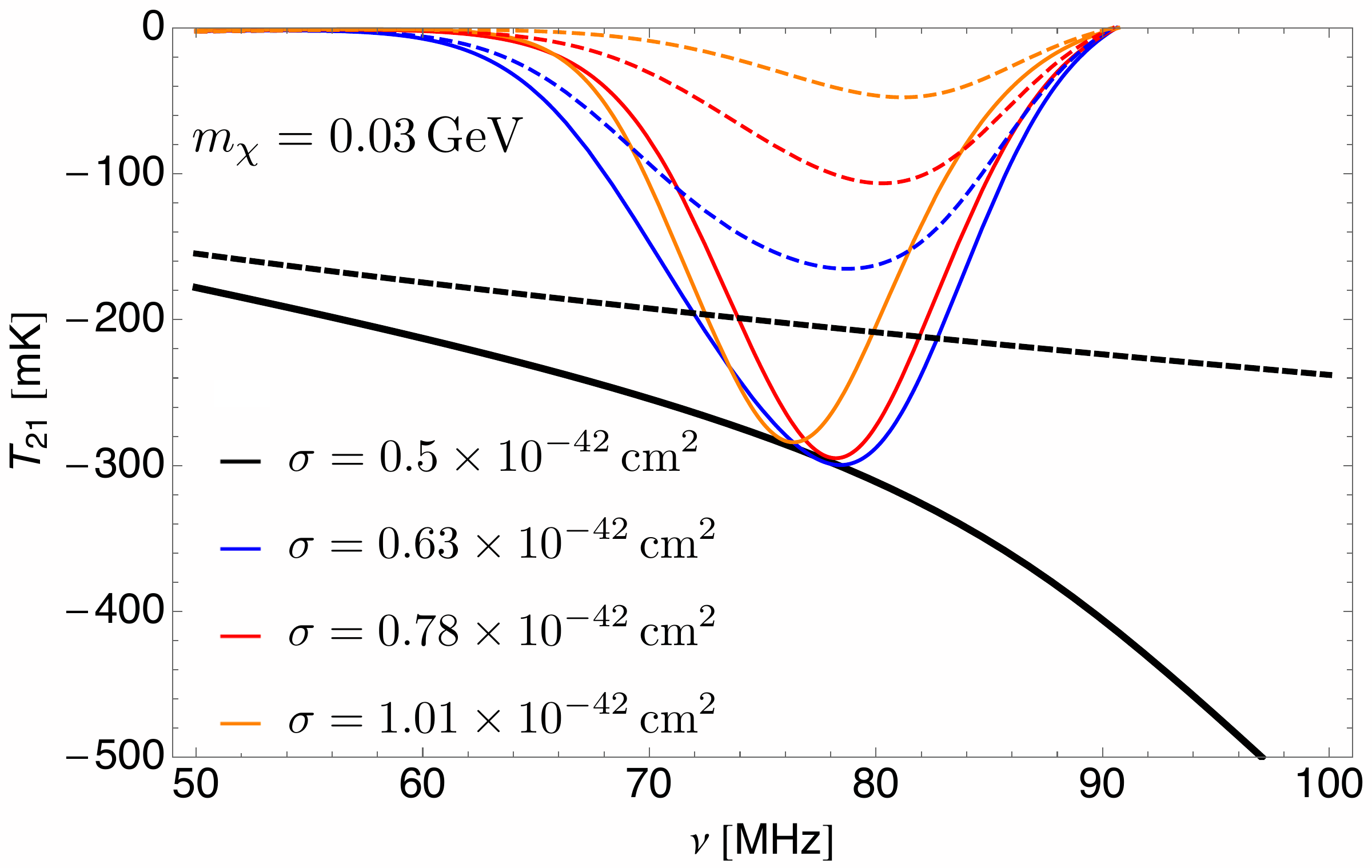}
\caption{The 21cm brightness temperature with (solid) and without (dashed) DM--baryon interactions for different choices of the Lyman-$\alpha$ coupling $x_\alpha$. Upper (Lower) panel: DM-particle mass is set to $0.3\,{\rm GeV}$ ($0.03\,{\rm GeV}$). Black lines correspond to infinite $x_\alpha$ (i.e.\ $T_s= T_b$); blue lines assume a rather large Lyman-$\alpha$ coupling, while red (orange) lines correspond to a fraction $20\%$ ($5\%$) of it (see Appendix~\ref{sec:appendix} for the full list of astrophysical parameters). As the efficiency of coupling the spin temperature to the gas temperature decreases, a much stronger cross section is needed to yield an absorption trough marginally consistent with the EDGES measurement.}
\label{fig:realSigma}
\end{figure}

While the door may appear to remain open (ever so slightly) to a DM interpretation of the EDGES signal, it is important to bear in mind that this minimum cross section heavily depends on the assumption regarding the Lyman-$\alpha$ coupling efficiency. This dependency is not linear and thus unintuitive.
To illustrate what a realistic signal would require, we plot in Figure~\ref{fig:realSigma} the 21cm absorption profile in the full frequency range measured by the EDGES low-band instruments, with and without DM-hydrogen interactions. In each case we set the scattering cross-section amplitude to the value necessary to roughly yield $\langle T_{\rm 21}(\nu=78.1\,{\rm MHz})\rangle=-300\,{\rm mK}$. We compare the most optimistic choice of setting $T_s= T_b|_{V_{\chi b,0}}$ (which was employed above to generate the 21cm curve in Figure~\ref{fig:CMBConstraints}), with a range of possible astrophysical scenarios for $T_s$. 

We use a simple phenomenological prescription---described in the Appendix---to model the astrophysical processes that take place during Cosmic Dawn, namely cosmic reionization, cosmic heating (which is usually attributed to X-ray emission from stellar remnants), and the Lyman-$\alpha$ coupling parameterized by $x_\alpha$ in Eq.~\eqref{eq:Ts}, \cite{Cohen:2016jbh}. Focusing on the latter, 
we consider three different values of $x_\alpha$, which under $\Lambda$CDM would yield a minimum temperature between $T_{21}\!=\!-40\,{\rm mK}$ and $T_{21}\!=\!-160\,{\rm mK}$. We set the reionization redshift and duration to values consistent with {\it Planck} 2015 cosmology and include a minimal amount of X-ray heating in the calculation of $T_b$, simply to ensure that the maximum absorption is reached near $\nu=78\,{\rm MHz}$ and the signal cuts off around $90\,{\rm MHz}$, roughly matching the EDGES measurement. 

Figure~\ref{fig:realSigma} demonstrates that in order to bring the absorption amplitude that would occur under $\Lambda$CDM (\textit{i.e.}, without DM--hydrogen interactions) to the minimal level that is marginally consistent with  EDGES, the cross section required can be significantly larger than the minimum cross section plotted in Figure~\ref{fig:CMBConstraints}, depending on the Lyman-$\alpha$ coupling and on the DM particle mass. We emphasize that we deliberately employ a very simple model for these quantities so as to render our conclusions most transparent. We have verified that the particular choices of the free astrophysical parameters, other than $x_\alpha$, have a minor effect on the 21cm absorption amplitude required for consistency with EDGES.

With data from future experiments such as CMB-S4~\cite{Abazajian:2016yjj}, an increasing portion of the allowed astrophysical parameter space to explain the EDGES signal with a direct $v^{-4}$ coupling of DM to hydrogen can be probed. Measurements of the global 21cm signal~\cite{Fialkov:2018xre} (also at higher frequencies), as well as the 21cm power spectrum~\cite{Cohen:2017xpx,Munoz:2018jwq,Kaurov:2018kez}, will complement the CMB constraints.

\subsection{Additional sources of heating}
To conclude this discussion, we list a number of heating sources, all of which, while uncertain to a degree, should make it even harder to cool the baryons enough to explain the EDGES result. For more details on these processes, we refer the reader to Ref.~\cite{Cohen:2016jbh} which charts the parameter space of relevant astrophysical models, and to Ref.~\cite{Venumadhav:2018uwn} which introduces a hitherto neglected heating effect, mediated by  Lyman-$\alpha$ emission from the first stars, on the baryon temperature.
Briefly, these effects include:
\begin{itemize}
\item An inevitable heating source comes from X-ray emission from  remnants of the first stars and subsequent generations. Depending on when this source becomes efficient, it could have a varying degree of influence on the maximum absorption possible. In the illustration we provided above, the duration of the X-ray heating phase was deliberately chosen to be quite minimal, $\Delta z_X=1$, and its central redshift set right before the end of the Cosmic Dawn epoch, $z_{X0}=12.75$, so as to yield an absorption profile shape that is consistent with the EDGES measurement, but at the same time remain conservative and incur only minimal heating of the baryons at the time of peak absorption. Stronger heating may require  higher scattering cross sections with DM in order to yield the measured EDGES trough.
\item An efficient Lyman-$\alpha$ coupling requires a strong Lyman-$\alpha$ flux, which in turn can result in additional, non-negligible heating of the baryon gas via a new mechanism very recently derived in Ref.~\cite{Venumadhav:2018uwn}. Lyman-$\alpha$ photons from the first stars were shown to mediate energy transfer between the CMB photons and the thermal motions of the hydrogen atoms, resulting in a $\mathcal{O}(10\%)$ alteration of the 21cm brightness temperature at $z\sim17$. For non-$\Lambda$CDM scenarios, such as DM--baryon scattering, the effect can be significantly stronger. A detailed treatment of this effect is beyond the scope of this work, but the preliminary analysis conducted in Ref.~\cite{Venumadhav:2018uwn} indicates that the 21cm absorption amplitude in the presence of cooling due to DM--baryon interactions can be halved when including this heating mechanism, which can require cross sections as much as an order-of-magnitude stronger to overcome it.
\item In addition to astrophysical sources of heating, in certain models which exhibit Coulomb-type interactions, annihilation of DM particles can lead to energy injection into the IGM and cause additional heating~\cite{Liu:2018uzy}. We plan to revisit this elsewhere~\cite{Dan}.
\end{itemize}
In view of this list, the illustration in Figure~\ref{fig:realSigma} should be considered quite conservative.

\section{Conclusions}
\label{sec:conclusions}

Non-gravitational interactions between  DM and standard model particles may lead to detectable imprints in cosmological observables such as the 21cm signal~\cite{Tashiro:2014tsa,Munoz:2015bca}. 
The Cosmic Dawn era provides a unique observational window to probe possible interactions between baryons and cold DM, as this is the point in time where the global gas temperature reaches its minimum value in the history of the Universe. For a $v^{-4}$ cross section, the interaction during this epoch is significantly enhanced versus earlier or later times, when the particle velocities are larger~\cite{Barkana:2018lgd}.  Such scattering was suggested as a possible explanation of the EDGES measurement of a stronger-than-expected 21cm absorption amplitude around redshift $z\sim17$.\footnote{Alternative explanations include a new source of radio emission in the EDGES band~\cite{Feng:2018rje,Fraser:2018acy, Ewall-Wice:2018bzf,Pospelov:2018kdh}; an earlier kinetic decoupling of baryons from CMB photons \cite{Hill:2018lfx,Falkowski:2018qdj,Poulin:2018dzj}; or foreground residuals~\cite{Hills:2018vyr}.}
  
However, this scenario may also lead to detectable signatures in other measurements,  in particular those of the small-scale CMB power spectrum~\cite{Dvorkin:2013cea,Xu:2018efh,Slatyer:2018aqg,paper1}. Still, as we demonstrated above, even for the very low cross-section amplitudes allowed by the CMB constraints from {\it Planck} 2015 data---which were derived in Ref.~\cite{paper1}---the effect on the 21cm signal can potentially be considerable.

Here we have investigated in greater detail two classes of models that have been suggested to explain the anomalously large 21cm absorption signal recently detected by the EDGES experiment. 
One is that of Ref.~\cite{Barkana:2018lgd}, which adopted the phenomenological approach in Ref.~\cite{Munoz:2015bca} and considered a direct interaction between the DM particles and the hydrogen gas, without a concrete particle physics model in mind. 
If DM interacts with neutral hydrogen, then weak cross sections---below the sensitivity of current CMB experiments---suffice in order to account for the EDGES signal. Barring a concrete model to explore in this case, we focused on the astrophysical uncertainties and illustrated that while it is possible for this phenomenological interaction to cool down the baryons to the desired level, the window for a realistic signal which incorporates all the sources of heating is at best limited. 

The more motivated model from the particle-physics perspective is millicharged DM. As it scatters only with ionized particles, the interaction is suppressed during Cosmic Dawn, when the ionization fraction is very low, $x_e\sim2\times10^{-4}$, and models with $100\%$ of DM in the form of millicharged particles require cross sections strongly ruled out by the CMB if they are to explain EDGES.
However, it has been claimed by Ref.~\cite{Munoz:2018pzp,Berlin:2018sjs,Barkana:2018qrx} that if millicharged DM comprises roughly $1\%$ of the total DM, it could explain the EDGES result while evading various astrophysical and collider constraints. 
 As this work  has shown, based on a detailed analysis of the effect of DM--baryon interactions on the 21cm signal, and taking into account constraints from the CMB, as well as stellar and SN1987A cooling, only a tiny window remains open for a DM-millicharge explanation of the EDGES anomaly. More explicitly, we found that for DM fractions $0.0115\%\left(m_\chi/{\rm MeV}\right)\lesssim f_{\chi}\lesssim 0.4\%$, particle masses $0.5\,{\rm MeV}-35\,{\rm MeV}$, and charge $10^{-6} e\lesssim\epsilon e\lesssim10^{-4} e$ (refer to Figure~\ref{fig:DMMillichargeParamSpace} for the precise range of viable charge fractions), the EDGES anomaly can in principle be explained (see our discussion of astrophysical uncertainties above).

Notably, in addition to explaining the EDGES measurement while evading CMB limits, a small fraction of DM which is tightly bound to baryons at recombination is a potentially interesting notion, as it could resolve the current tension between CMB and BBN measurements of the baryon energy density. While this component would appear as excess baryon energy density in the former, it would be missing from the latter, which is directly based on the deuterium abundance (and also arises from an earlier epoch where the interacting DM and the baryons may not yet be strongly coupled). Comparing the two values, $\Omega_b h^2=0.0224\pm0.0001$  ({\it Planck} 2018 \cite{Aghanim:2018eyx}) vs. $\Omega_b h^2\simeq0.02170\pm0.00026$ (BBN, taking the average and larger uncertainty between the values found in two independent measurements~\cite{Cooke:2017cwo,Zavarygin:2018dbk}), they are discrepant at a $\gtrsim2\%$ level, which indeed could be alleviated if the fraction of this DM component is $f_\chi\!\sim\!0.4\%$. Importantly, CMB-S4 will probe down to $f_\chi\!\sim\!0.1\%$~\cite{paper1,dePutter:2018xte}, enabling a direct test of the DM explanation for the $\Omega_b$ discrepancy with BBN. (For alternative solutions, see Ref.~\cite{Cooke:2017cwo}.)

In conclusion, the DM interpretation of the EDGES anomaly (which awaits corroboration by other global 21cm experiments~\cite{Voytek:2013nua,Bernardi:2016pva,Singh:2017gtp}), currently hangs on a thin thread. Fortunately, future experiments such as CMB-S4, as well as experiments targeting the Lyman-$\alpha$ forest power spectrum~\cite{Dvorkin:2013cea, paper1, McDonald:2004xn} and the power spectrum of 21cm fluctuations~\cite{Koopmans:2015sua,DeBoer:2016tnn,Patil:2017zqk,Chatterjee:2018vpw}, will be able to probe these scenarios directly and may provide a definitive answer.

\acknowledgments

It is our pleasure to thank Ilias Cholis, Dan Pfeffer and especially Julian Mu\~noz for help and useful discussions. EK is grateful for the hospitality of LITP at the Technion, Israel, while 
KB and VG acknowledge KITP and {\it The Small-Scale Structure of Cold(?)~Dark Matter} workshop for their hospitality and support under NSF grant PHY-1748958, during the completion of this work. 
This work was supported at Johns Hopkins University in part by NSF Grant No.~1519353 and by NASA Grant
NNX17AK38G, and the Simons Foundation. VG gratefully acknowledges the support of the Eric Schmidt fellowship at the Institute for Advanced Study. For RB, this publication was made possible through the support of a grant from the John Templeton Foundation; the opinions expressed in this publication are those of the author and
do not necessarily reflect the views of the John Templeton Foundation. RB was also supported by the ISF-NSFC joint research program (grant No.\ 2580/17).
Part of this work has been done thanks to the facilities offered by the Universit\'e Savoie Mont Blanc MUST computing center.

\appendix

\section{Initial Conditions for 21cm Calculation}

To generate initial conditions for solving Eqs.~\eqref{eq:V}-\eqref{eq:Tb} from a computation that appropriately handles the pre-recombination physics, we used the method developed in Ref.~\cite{paper1}. There, we 
modified \texttt{CLASS} to incorporate scattering between baryons and DM. For the CMB calculation in Ref.~\cite{paper1}, the value of $V_{\chi b}$ could be safely taken as the square root of its variance to obtain the average temperature evolution. Since the baryons are tightly coupled to the photons, the effect of DM--baryon scattering on the baryon temperature is negligible prior to recombination.

Here, in order to take into account the fact that $V_{\chi b}$, $T_\chi$, and $T_b$ vary between different patches in the sky and properly calculate the average 21cm brightness temperature, Eq.~\eqref{eq:T21V}, we sampled a range of initial relative velocities $V_{{\chi b},0}$ at $z=1700$ (with root-mean-square $\langle V_{\chi b}^2 \rangle$). Then, assuming that each patch of sky had a constant $V_{\chi b}$ from very early times to the epoch of recombination, we kept the velocity in the patch constant and evolved the DM and baryon temperatures $T_\chi$ and $T_b$ to the same redshift. From that redshift on we used Eqs.~\eqref{eq:V}-\eqref{eq:Tb} to track the remaining evolution until Cosmic Dawn. We chose to take our initial conditions at $z=1700$, before $V_{\chi b}$ is expected to substantially decrease as the baryons decouple from the photons. This procedure neglects the effect of baryon-photon scattering on the relative velocity, which is not included in Eq.~\eqref{eq:V}. 
To check this approximation, we also considered an alternative scheme, whereby the initial conditions were taken at a lower redshift of $z=800$, at which we averaged over $V_{\chi b}$ values (sampled with root-mean-square $\langle V_{\chi b}^2 \rangle$ taken from the Ref.~\cite{paper1} computation) but used the average temperature values (given by the same code), and obtained  similar results.

\section{21cm Cosmic Dawn  Modeling}
\label{sec:appendix}
Consistent with the well-known ``turning points" model for the 21cm brightness temperature evolution at high redshift~\cite{Pritchard:2011xb}, our model for the 21cm signal at Cosmic Dawn accounts very simply for the following effects:
\begin{itemize}
\item The Lyman-$\alpha$ radiation from the first starts couples the spin temperature to the gas temperature (the Wouthuysen-Field effect), through the Lyman-$\alpha$ coupling coefficient $x_\alpha$ in the expression for the spin temperature $T_s$ in Eq.~\eqref{eq:Ts}.
\item X-ray heating from stellar remnants is included in the evolution of the gas temperature $T_b$, by adding a corresponding term to Eq.~\eqref{eq:Tb}.
\item The radiation from the first stars gradually ionizes the gas and adds to the ionization fraction $\bar{x}_e$ in Eq.~\eqref{eq:xe}, which is used in Eq.~\eqref{eq:T21}.
\end{itemize}

Motivated by Refs.~\cite{Mirocha:2015jra,Harker:2015uma}, we use a {\it tanh} parameterization for the Lyman-$\alpha$ coupling coefficient, the ionization fraction, and the X-ray heating contribution:
\begin{equation}
  A_{\mathrm{i}}(z)=A_{\mathrm{i}}\left(1+\tanh[(z_{\mathrm{i}0}-z)/\Delta z_{\mathrm{i}}]\right)\ , 
  \label{eq:tanh}
\end{equation}
where the nine free parameters in this model are simply the step height $A_{\mathrm{i}}$ ($i$ stands for either ``$\alpha$", ``xe" or ``X''), pivot redshift $z_{i0}$, and duration $\Delta z_i$ for each quantity.
In practice, we hold eight of these parameters fixed and vary {\it only} the Lyman-$\alpha$ coupling amplitude $A_{\alpha}$.
To approximately reproduce the EDGES absorption profile, we use the following set of fiducial values: 
\begin{align}
&\{A_{\alpha}, z_{\alpha0}, \Delta z_\alpha\}=\{100,17,2\} \nonumber \\
&\{A_{{\rm xe}}, z_{{\rm xe}0}, \Delta z_{\rm xe}\}=\{1,9,3\} \nonumber \\  
&\{A_{X}, z_{X0}, \Delta z_X\}=\{1000\,{\rm K},12.75,1\} \nonumber
\end{align}
We emphasize that the particular choice of values has no bearing on the conclusions of our investigation in Section \ref{sec:Astrophysics}.
Following Ref.~\cite{Mirocha:2015jra}, we define $x_{\alpha} \equiv 2 A_{\alpha}(z)/ (1 + z)$, add $d A_X(a)/da$ to Eq.~\eqref{eq:Tb} and $A_{\rm xe}(z)$ to Eq.~\eqref{eq:xe}. For the illustration in Figure~\ref{fig:realSigma}, we vary $A_{\alpha}$ as a fraction of its fiducial value and consider $20\%$ and $5\%$ cases (i.e.\ $A_\alpha=20$ and $A_\alpha=5$). 

Lastly, we set the collisional coupling coefficient $x_c$ according to the approximation in Ref.~\cite{Pritchard:2011xb}, i.e.\ $x_c=n_H\kappa_{10}/A_{10}$, where $A_{10}$ is the Einstein coefficient, $n_H(z)$ is the hydrogen number density, and $\kappa_{10}$---the specific rate coefficient for spin de-excitation by hydrogen atom collisions---depends on the gas temperature at each redshift and is well-approximated in the relevant range of temperatures by $\kappa_{10}=3.1\times10^{-11}T_b^{0.357}e^{-32/T_b}$.

\end{document}